\documentclass[letterpaper]{jpconf}
\usepackage{graphicx}
\begin{document}
\title{Recent $\nu$s from IceCube}

\author{Spencer R. Klein, for the IceCube Collaboration}

\address{Nuclear Science Division, Lawrence Berkeley National Laboratory, Berkeley, CA, 94720 USA}

\ead{srklein@lbl.gov}

\begin{abstract}

IceCube is a 1 km$^3$ neutrino detector now being built at the South Pole.  Its 4800
optical modules will detect Cherenkov radiation from charged particles produced in neutrino interactions.
IceCube will search for neutrinos of astrophysical origin, with energies from 100 GeV up 
to $10^{19}$ eV.  It will be able to separate $\nu_e$, $\nu_\mu$ and $\nu_\tau$.   In addition to
detecting astrophysical neutrinos, IceCube will also search for neutrinos from WIMP annihilation in the
Sun and the Earth, look for low-energy (10 MeV) neutrinos from supernovae, and search for a host of
exotic signatures.  With the associated IceTop surface air shower array, it will study cosmic-ray
air showers.  

IceCube construction is now 50\% complete.  After presenting preliminary results from the partial detector,
I will discuss IceCube's future plans. 

\end{abstract}

\section{Introduction}

Despite decades of research, the sites for acceleration of high energy cosmic rays have
not been found.   Because protons and heavier nuclei are bent in interstellar and intergalactic
magnetic fields, the trajectories of detected cosmic-rays do not point back to
their sources.  Some sources of TeV photons have been observed, but these photons
may be from inverse Compton scattering, and thus be
evidence of electron acceleration rather than hadron acceleration.

Neutrinos are produced
in charged pion decays; these pions may be produced in collisions between the cosmic
rays under acceleration and either photons or gas at the acceleration site (`beam-gas
collisions').  Neutrinos have small interaction cross sections and so escape from 
dense sources, and are undeflected by magnetic fields, so they are attractive probes.  

Because of the small cross sections, a very large detector is
required to observe neutrino signals \cite{PDD}.  Two different calculations, one based on the
measured cosmic-ray flux, and a second based on observed TeV $\gamma$-ray signals, find that
a detector with a volume of 1 km$^3$ should observe neutrino signals.  This formed the rationale for
IceCube \cite{PT}.

\section{Detector Design}

\begin{figure}[tp]
\center{\includegraphics[clip,scale=0.40]{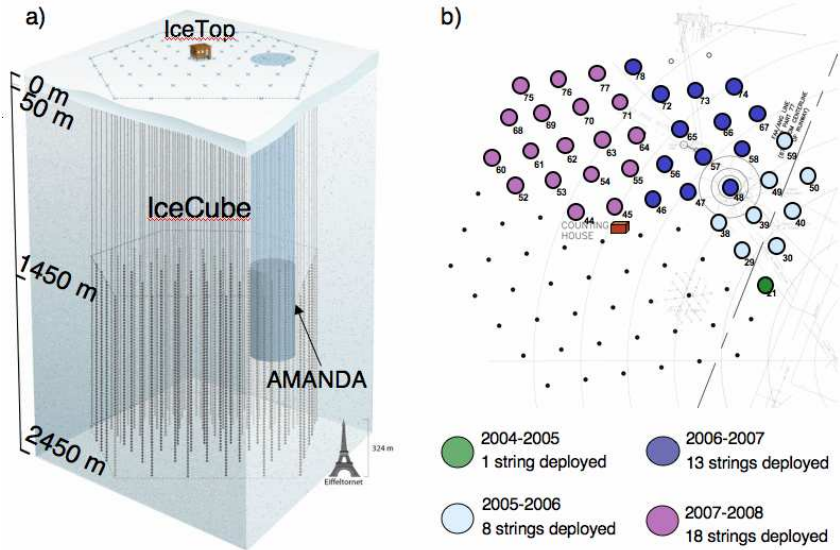}}
\caption[]{(Left) A schematic view of the IceCube detector. The darker region shows the location
of the smaller AMANDA array that preceded IceCube. (Right) The IceCube layout, showing 
construction progress.  
\label{fig:overview}}
\end{figure}

The baseline IceCube design is shown in Fig. \ref{fig:overview} \cite{PDD}.
Sensors are deployed on 80 vertical strings, each consisting of 60 digital optical modules (DOMs)
attached to a 2500 m long cable.  The DOM spacing is 17 m; the instrumented region 
is between 1450 m and 2450 m below the surface.
The strings are placed on a hexagonal grid with 125 m spacing, covering 1 km$^2$. The associated
IceTop surface air shower array is made up of
160 ice-filled tanks, two near the top of each string.  
%Each tank is instrumented with two DOMs.  
Two DOMs in each tank detect Cherenkov radiation from charged particles in air showers.

Each DOM collects data autonomously.  A 35 cm diameter
pressure vessel (Benthosphere) holds a 25 cm diameter Hamamatus R7081-02 photomultiplier tube (PMT), plus associated
electronics.  The electronics package includes a PMT base, Cockroft-Walton high voltage generator, resistive
divider PMT base, flasher board (containing 12 light emitting diodes, with programmable drivers), and
a ``Main Board" (MB) containing a complete data acquisition (DAQ) system \cite{SORMA}.  The DAQ
includes two separate waveform digitizer systems.  One uses a custom switched-capacitor array
chip to collect 128 samples of the PMT output at 300 megasamples per second (MSPS); three independent channels 
for each PMT provide 14 bits of dynamic range.  The second uses a commercial 40 MSPS 10-bit ADC chip; it
records 6.4 $\mu$s of data after each trigger.  

Data acquisition is launched by a discriminator
with a threshold set at about 0.35 of a photoelectron.   A local coincidence
circuit regulates data collection.  If a DOM's nearest or next-to-nearest neighbor on a string fires
within 1 $\mu$s of the discriminator crossing, then full waveform data is transmitted to the surface. 
A future software update will save some information from isolated hits: the DOM
launch time, plus the 3 highest samples from the 40 MSPS ADC.  All of the data is compressed, 
assembled into packets, and transmitted to the surface.  

The DOMs have internal self-calibration circuitry.  Each DOM has a
precision clock.  Every 3 1/2 s these clocks are recalibrated to a master clock on the surface.  This is done
by sending a signal up and down the cable.  The time resolution is about 2 ns, across the entire detector
\cite{Firstyear}. On-DOM LEDs allow for automatic amplitude/PMT gain calibration \cite{ICRC}.

The DOMs are connected to the surface electronics via a cable which contains 30 twisted pair lines
(each pair serves 2 DOMs), plus a strength member.   Each twisted pair carries $\pm 48$ V DC 
which powers the DOM, plus
digital data and timing calibration signals.  

Because of the harsh environment, the DOMs must meet stringent requirements regarding power consumption
(3.5 W/DOM), temperature range (to $-55^0$C), and reliability. About 98\% of the DOMs are working to full specifications, and
another 1\% are usable, but impaired in some way.  The most common problem is in the local coincidence connections. 

\section{IceCube Construction}

The holes for the IceCube strings are drilled using a hot-water drill.  A 5 MW heating system
produces 760 liters/minute of 88$^0$ C water.  This water is propelled through a 1.8 cm
diameter nozzle at 200 pounds/square inch, melting a hole through the ice.  It takes about 40 hours
to drill each 2500 m deep, 60 cm diameter hole.  Once the hole is drilled, it takes about 12 hours to
lower the string of DOMs into the hole. 

Because of the Antarctic weather, the construction season is short, from November to mid-February.
Logistics is a key concern.  All of the personnel and hardware must be flown in on ski-equipped LC-130 
transport planes. 
The right panel of Fig. \ref{fig:overview} shows how construction has progressed since the first string
was deployed in early 2005.  By the end of the 2007/8 construction season, 40 strings were deployed.  

\section{Triggers and Data Collection}

IceCube has several triggers, all implemented in software. These triggers use the times
for the hits on all DOMs.   A multiplicity trigger selects time
windows when 8 DOMs fired within 5 $\mu$s.  In 2008, this was supplemented with a string trigger
which produced a trigger if 5 out of 7 adjacent DOMs fired within 1.5 $\mu$s; this increases our
sensitivity to low energy neutrinos.  A topological trigger is also under consideration; this would
be optimized for low energy, roughly horizontal neutrinos.
When a trigger condition is satisfied, the DAQ system records all hits within a $\pm 10$
$\mu$s time window.  When time windows from multiple triggers overlap, they are combined into a single window.  

Events selected by the triggers are fed to a set of software filters, which perform a variety of simple
reconstructions.  Filters look for upward going muons, cascades ($\nu_e$, $\nu_\tau$ and neutral
current interactions), contained events of any sort, extremely high energy events, events where a track 
starts or stops in the detector, and air showers.
Another filter selects events that come from near the position of the moon, to look for the moon shadow.
Currently, about 6\% of the events (80 Hz) pass at least one of these criteria.  These events are
transferred to the Northern hemisphere over a satellite link, allowing for rapid analysis.
The satellite bandwidth is about 32 Gbytes/day.  All of the data are stored on tape \cite{SORMA}. 

Table \ref{tab:datasets} shows the different IceCube datasets.  The
neutrino rate rises from about 1.5 $\nu$/day with 9 strings to a projected 200 $\nu$/day with 80 strings.
The rate increases faster than the detector size because of edge effects. 

\begin{table}
\center{
\begin{tabular}{lcrrrr}
Configuration & Year & Run & CR $\mu$ & $\nu$ Rate & Trigger \\
 (\# Strings)          &      & Length & Rate &    & Rate \\
\hline
1 & 2005 & -     & -     & 2 (total)    & -     \\
IC-9 & 2006 & 137 d & 80 Hz & 1.5/d & 150 Hz \\ 
IC-22& 2007 & 319 d & 550 Hz&20/d   & 670 Hz \\
IC-40& 2008 & 1 y   & 1000 Hz & -   & 1400 Hz\\
IC-80& 2011 & 10 y  & 1650 Hz & 200/d & TBD  \\
\end{tabular}
}
\caption{
\label{tab:datasets}
The datasets collected with IceCube at different stages of construction.}
\end{table}

\section{Reconstruction and Performance}

\begin{figure}[t]
\begin{minipage}{3 in}
\center{\includegraphics[clip,scale=0.54]{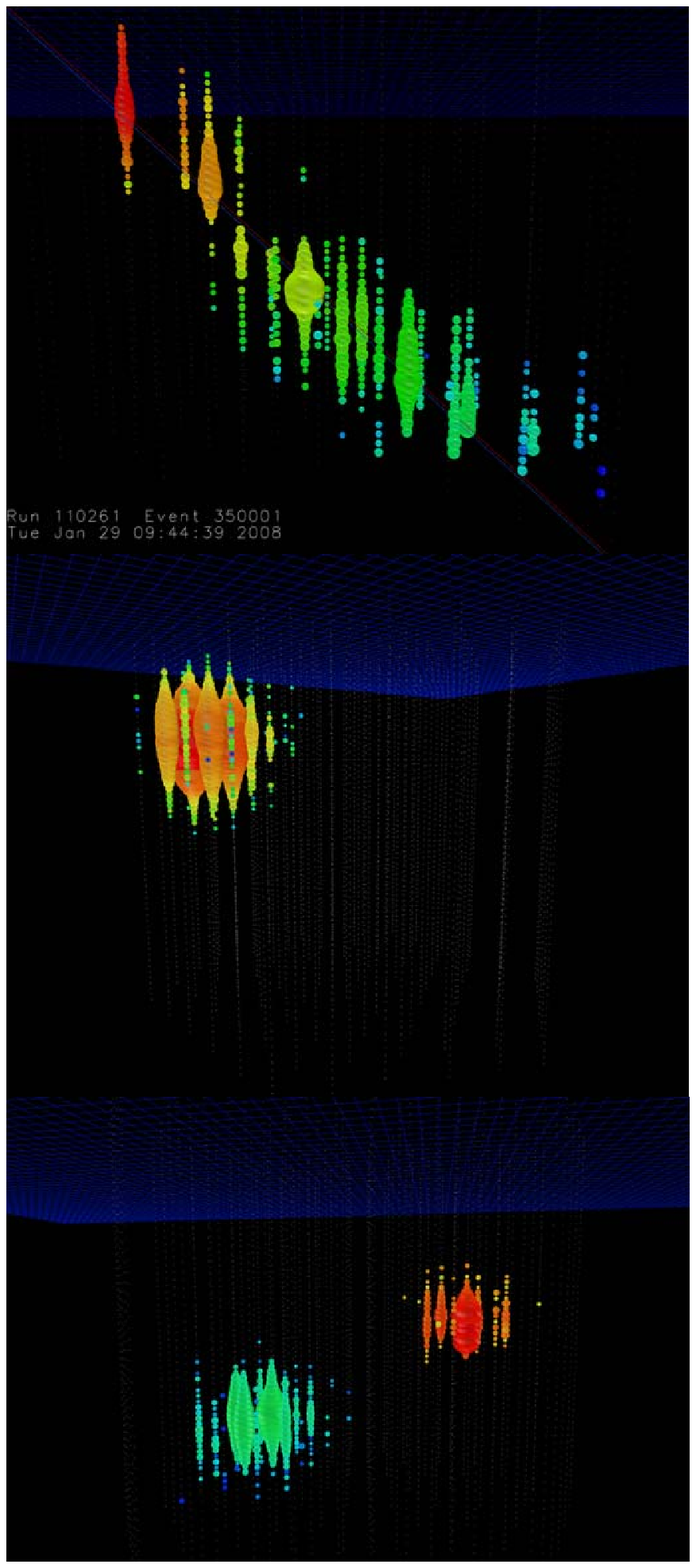}}
\caption{\label{fig:flavors}Event displays for (top) an actual muon (or muon bundle) in IC-40,
(middle) simulated $\nu_e$ event, and (bottom) simulated double-bang $\nu_\tau$ event.  The colors
indicate times, from red (earliest) to blue (latest).}
\end{minipage}\hspace{2pc}%
\begin{minipage}{3 in}
\center{\includegraphics[clip,scale=0.59]{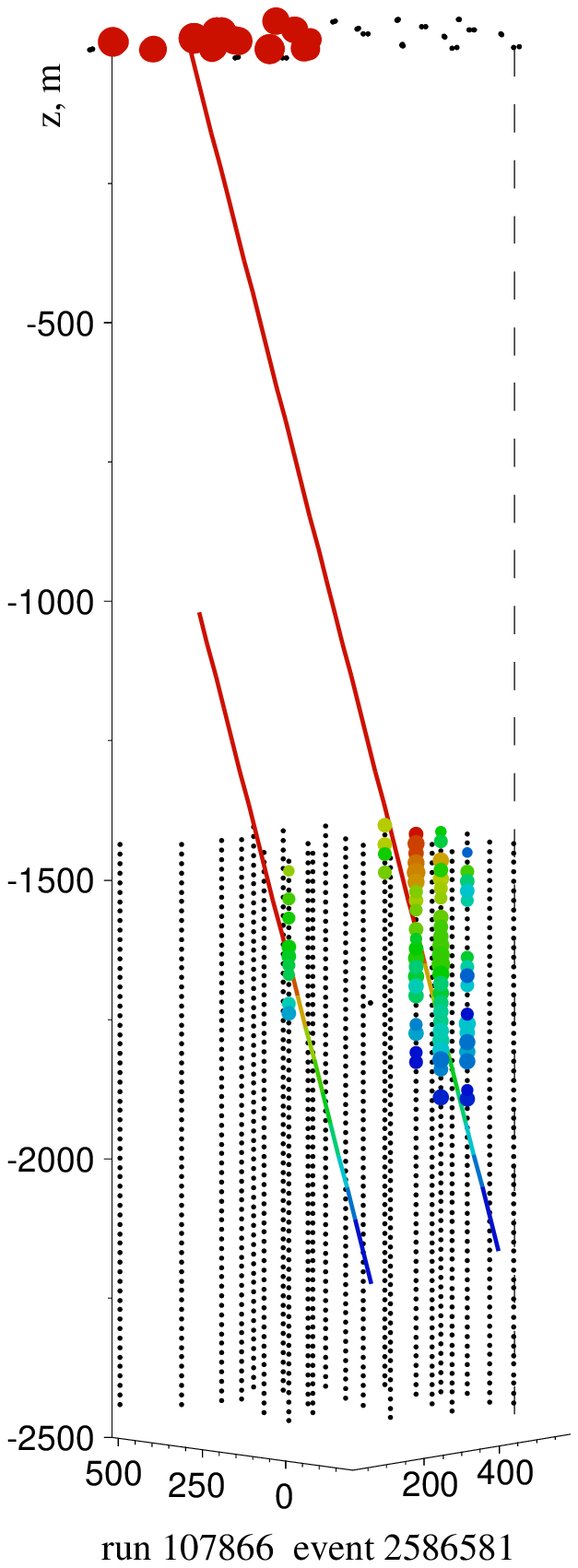}}
\caption{\label{fig:highpt}An IceCube-IceTop coincident event.  An air shower hit 11 IceTop tanks, while
the accompanying muon bundle caused hits in 84 DOMs in four strings; an additional muon 
about 400 m from the shower core was seen by 12 DOMs in another string \cite{ICRC}. 
}
\end{minipage} 
\end{figure}

IceCube is designed to identify and reconstruct all three flavors of neutrinos:
$\nu_\mu$, $\nu_e$ and $\nu_\tau$.  Figure \ref{fig:flavors} shows examples of these 
three topologies; the muon event is data, but the
cascade and tau are simulations.   IceCube events are reconstructed using algorithms 
designed to select these events
based on their different topologies: long tracks from muons, blob-like cascade events, 
$\tau$ double-bang events, etc.

Here, we focus on muon reconstruction \cite{muonreco}.  Cherenkov photons are produced by the muon and 
particles produced by muon interactions; for high-energy muons (above 1 TeV), the associated particles
produce most of the light.  These photons are emitted at the Cherenkov angle, about 41$^0$ in ice.  
In contrast to a typical charged particle detector, these
photons are observed at a significant (up to 30 m) perpendicular distance from the radiating particles.
Photon scattering in the ice affects the photon trajectory. 
Instead of fitting to points on a track, IceCube reconstruction algorithms must account for both the
distance travelled by the photon and its travel time. 

Muons are initially reconstructed with a first-guess
algorithm which fits the photon arrival times to a moving plane; the muon direction is perpendicular to the
plane.  Later reconstruction methods use maximum-likelihood fits.  These fits use probability distribution
functions which account for photon scattering and absorption.  These functions give the arrival time
distributions for photons from an infinite linear track to reach a DOM, as a function of the perpendicular distance
and angles.   The functions are depth dependent to account for the varying optical properties of the ice.
Noise is also included in the functions.  The track angular resolution depends on the visible length of the track 
(the lever arm); the median resolution will be better than $1^0$ for IC-80.   

\begin{figure}[bt]
\begin{minipage}{3 in}
\center{\includegraphics[clip,scale=0.40]{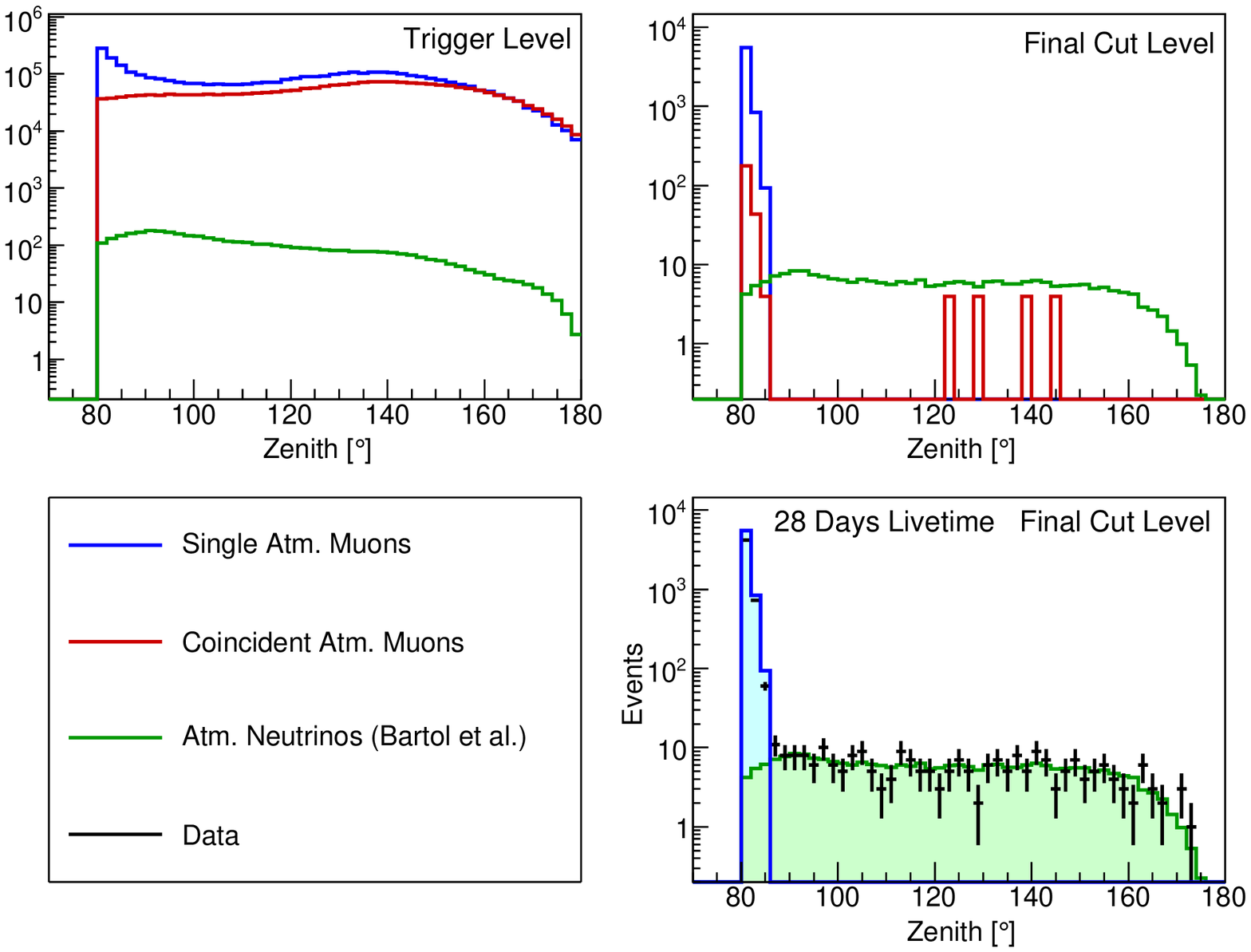}}
\caption{\label{fig:neutrinoangles} Preliminary expected neutrino signals and backgrounds, before and after cuts,
compared with data.}
\end{minipage}\hspace{2pc}%
\begin{minipage}{3 in}
\center{\includegraphics[clip,scale=0.48]{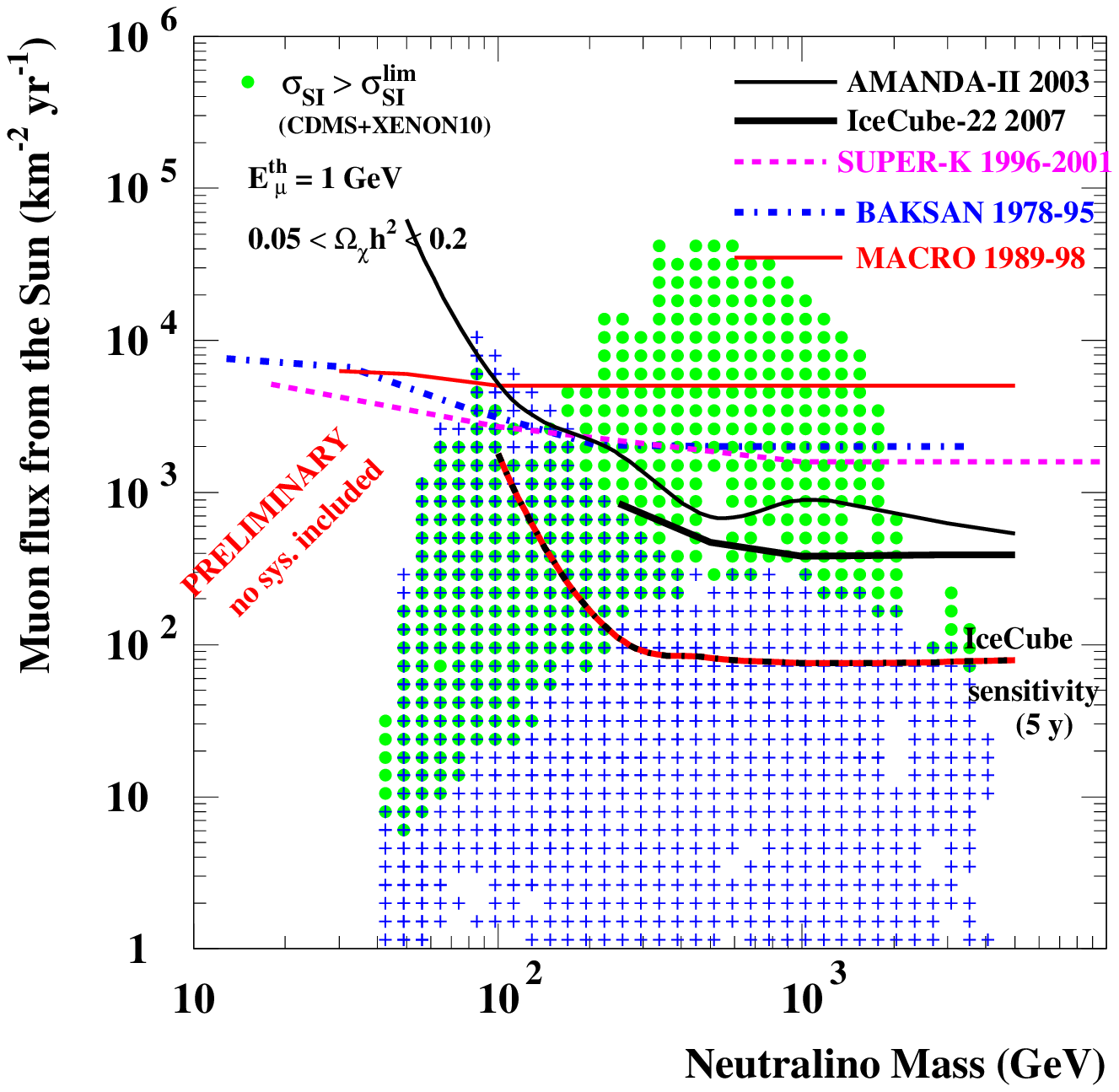}}
\caption{Preliminary IC-22 limits on neutrinos from WIMP annihilation in the Sun.
\label{fig:WIMPS}
}
\end{minipage} 
\end{figure}

\section{Results}
 
IceCube studies a wide variety of physics topics.  Besides searches for point sources of $\nu_\mu$, 
IceCube is also searching 
for diffuse extra-terrestrial $\nu_\mu$, $\nu_e$ and $\nu_\tau$.  Other physics topics include 
searches for $\nu$ from Gamma-Ray Bursts (GRBs) and from Weakly Interacting Massive Particle (WIMP) annihilation 
in the Earth 
or the Sun, studies of atmospheric $\nu$, searches for MeV $\nu$ from supernova collapse, and searches for a 
variety of exotica.  These exotica include magnetic monopoles, strangelets, and pairs of upward going particles.
The latter are produced in some models of supersymmetry with high mass scales, and also in some Kaluza-Klein theories. 

In all of the neutrino studies, the backgrounds are large.
The ratio of downgoing cosmic-ray muons to upgoing muons from neutrinos is about 500,000:1, so stringent
cuts are required to eliminate background.   
Cuts are applied to the reconstructed zenith angle (using multiple reconstruction algorithms), likelihood
of the fit, and the estimated angular resolution of the individual event (this is an indicator of quality).
These cuts remove most of the downward going muon bundles.  However,
IceCube is large enough that overlapping, independent cosmic-ray muons within a single trigger window constitute 
a significant background; additional cuts are needed to remove these events.
Figure \ref{fig:neutrinoangles} shows the zenith angle distributions of the signals and backgrounds.
The data and simulations are in good agreement; after cuts, the
upward-going events are dominated by atmospheric neutrinos.

\begin{figure}[bt]
\center{\includegraphics[clip,scale=0.40]{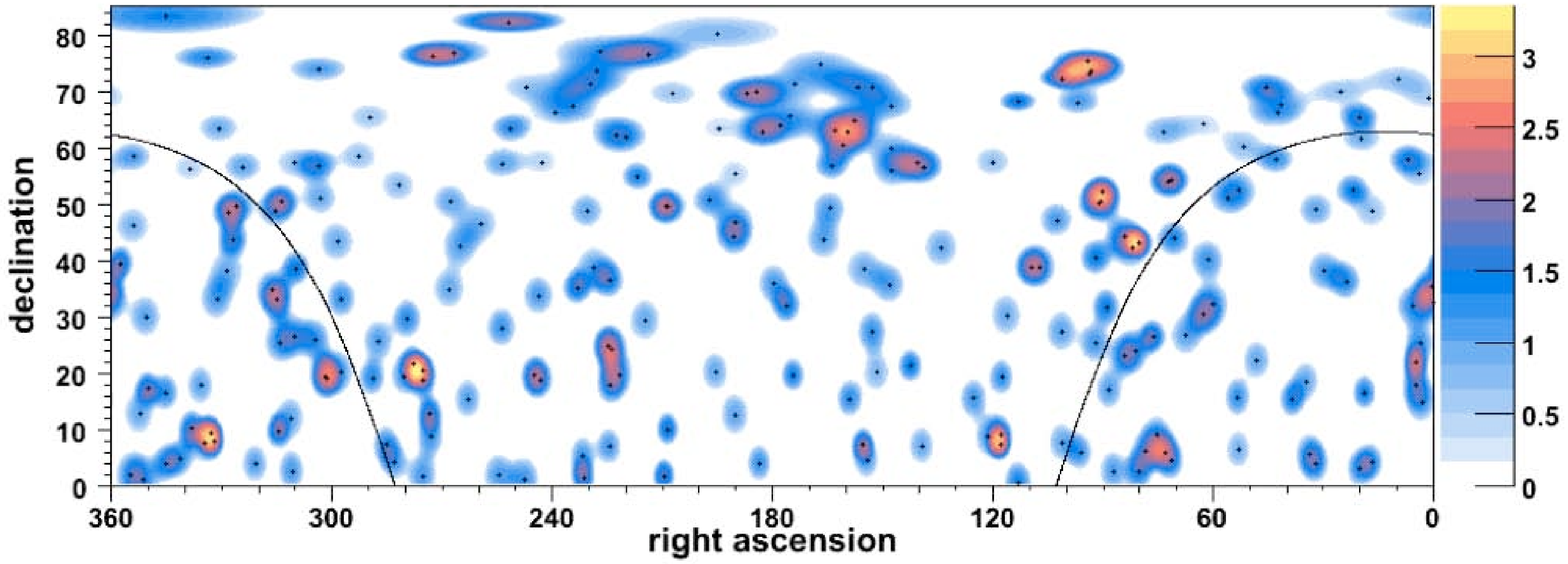}}
\vskip -0.25 in
\center{\includegraphics[clip,scale=0.57]{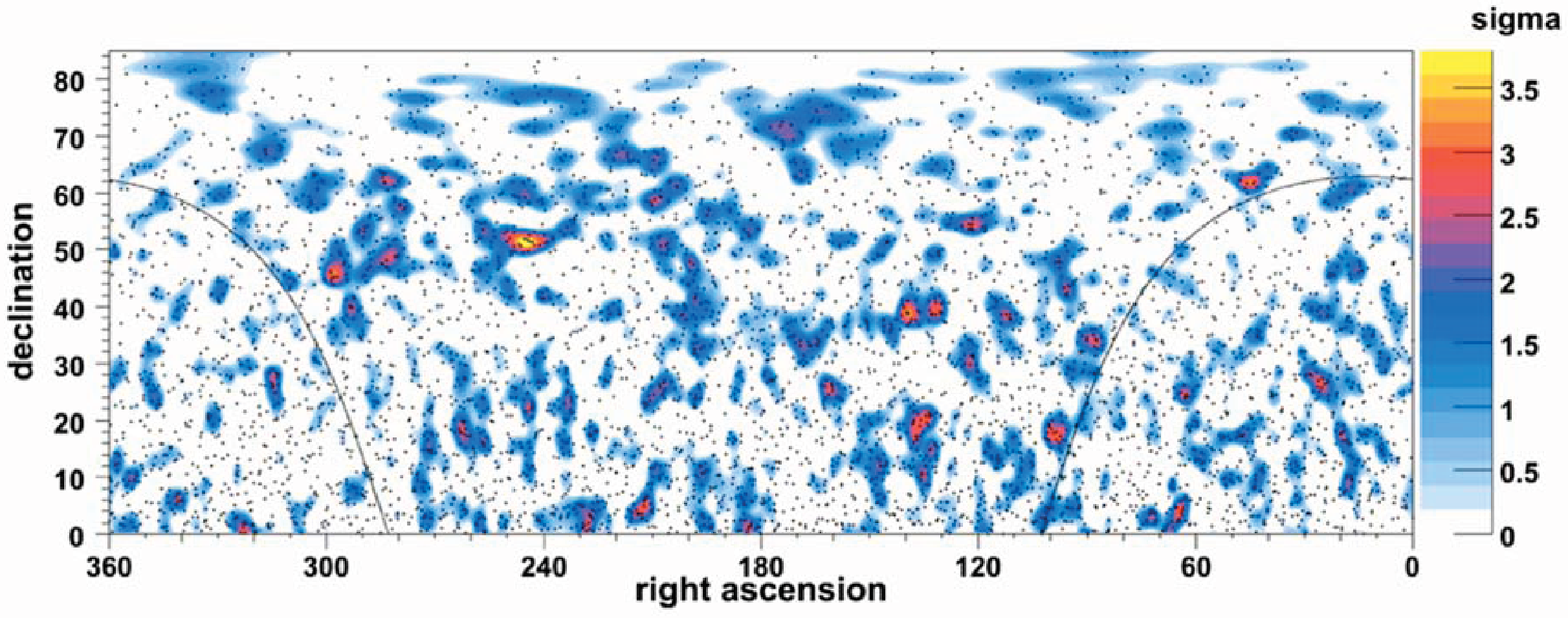}}
\caption[]{(Top) Neutrino sky map from IC-9. (Bottom) Scrambled neutrino sky map from IC-22.
\label{fig:skymap}}
\end{figure}

The arrival directions of these neutrinos can be searched for clusters of events from
neutrino sources. 
The top panel of Fig. \ref{fig:skymap} shows the sky-map produced from the 9-string data \cite{9string}.  
It contains 234 events (with about 90\% $\nu$ purity) collected over 137 d.  No sources were seen, in either a 
search for $\nu$ from 26 specific possible sources, or in an all-sky search.  The average
sensitivity for the all-sky search, assuming an $E^{-2}$ spectrum, is $d\phi/dE = 1.2\times10^{-10}$
TeV$^{-1}$ cm$^{-2}$ s$^{-1}$ (E/TeV)$^{-2}$.  
The bottom panel of Fig. \ref{fig:skymap} shows a scrambled IC-22 sky map, containing
about 5,000 events collected during 250 d (20 $\nu$/d).  These events have been scrambled in
right ascension, but are representative of what we expect after unblinding.
With a typical resolution of 1.5$^0$, the limits from this search should be about 5 times better than for IC-9.   

Searches for diffuse $\nu_\mu$ \cite{Gary} and $\nu_e$ must rely on the expected harder energy spectrum of 
extra-terrestrial
$\nu$, compared to atmospheric $\nu$. Extra-terrestrial $\nu$ are expected to have an
$E_\nu^{-2}$ spectrum, while the atmospheric $\nu$ spectrum is much softer, about $E_\nu^{-3.7}$.

IceCube also studies transient $\nu$ sources.  Gamma-Ray Bursts (GRBs) are perhaps the
most prominent. During the recent extremely bright burst, GRB080319B, IceCube was running in a 9-string
test mode.  A fireball model with a Lorentz boost of 300 predicts that we should see about 0.1 event; an
analysis is underway. 

If dark matter is composed of weakly interacting massive particles (WIMPS), these WIMPs may
be gravitationally captured in the earth and the sun.  They may then annihilate, producing a neutrino
signal.  
Figure \ref{fig:WIMPS} shows the limits from a search for $\nu$ from the Sun in IC-22.
The search used about 4 months of data from the period when the sun was below the horizon at the South Pole. 
No signal was found, and limits have been set on WIMPs with masses  between 250 GeV and 5 TeV.  

The IceTop surface array is being used to measure the cosmic-ray energy spectrum and study its
composition. At energies above several PeV (above the knee), where the efficiency is high, the
measured spectrum agrees well with the spectral index of 3.05 found by other experiments \cite{tilo}. 
The high-energy cosmic-ray flux is nearly isotropic, but
the dependence on observed zenith angle depends on composition;
at large zenith angles the pathlength in the atmosphere is long and
showers from heavier nuclei are attenuated more than those from protons. 
Preliminary analysis of observed angular dependence
favors a mixed composition~\cite{tilo}.  The ratio of muon signal in
deep IceCube to shower size in IceTop in events seen
in both parts of IceCube will give a complementary measure of primary
composition. 

IceCube will measure muon decoherence at distances far from the shower core.  Muon pairs with 
separations of more than 100-200 m should be observable in IceCube; Fig. \ref{fig:highpt} shows one
air shower event with a muon bundle near the core, plus an isolated track about 400 m away.
The techniques used to reconstruct multiple $\mu$ in these events can also be used
to search for pairs of upward going charged particles.  These pairs are expected in some new-physics models,
most notably supersymmetry if the mass scale is high \cite{SUSY}.

Although the IceTop array has a trigger threshold of about 300 TeV, individual tanks are sensitive to
single particles from cosmic rays with much lower energies. A burst of particles with energies
of a few GeV will increase the counting rate in the tanks.   So, the tanks are sensitive to
solar flares which produce Ground Level Enhancements (GLE).
Because of the large overall tank area, IceTop is a sensitive solar activity monitor.
For example, a December 13, 2006 solar flare produced a 1\% increase in average tank counting rate in
a roughly 30 minute period.  Simultaneous increases were observed by
solar neutron monitoring stations spread around the globe.
The rate increases depend on
the individual tank thresholds, offering the possibility for inferring the energy spectrum of the 
incident particles. 

\section{Future Plans}

IceCube is scheduled for completion in 2011.   In addition to the baseline
80 strings, we are also developing a ``Deep Core" infill array which will 
have greatly increased sensitivity to low energy neutrinos.  \cite{Ty}.
Deep Core will consist of 6 additional strings with a 72 m grid spacing, in the
center of IceCube.  Its
DOMs will use new high quantum-efficiency phototubes which will be spaced
every 7 m in the deepest, clearest 350 m of ice. The rest of IceCube will 
serve as a veto for Deep Core.

We are also reviewing the locations of the final IceCube strings.  If some of the
outermost strings were moved further from the center of IceCube, our effective volume for
high-energy (PeV and EeV) neutrinos would be enhanced \cite{Dima}.

Also under consideration are radio and acoustic detectors optimized for EHE neutrinos \cite{Rolf}.  

\section{Conclusions}

The IceCube detector is currently 50\% complete.  The hardware is working extremely well, and we
have presented results on searches for point sources of neutrinos and neutrinos from the Sun.
IceTop has made a preliminary measurement of the cosmic-ray energy spectrum.  
Construction will be complete in 2011; the final detector will comprise the ``Deep Core"
infill array in addition to the baseline detector. 

It is a pleasure to acknowledge support from the U.S. National Science Foundation and the 
Department of Energy.

\smallskip

\end{document}